\documentclass[preprint]{aastex}
\usepackage{graphicx}
\received{2005 January 3}
\begin{document}

\title{REDSHIFTS IN THE SOUTHERN ABELL REDSHIFT SURVEY CLUSTERS. I. THE DATA}

\author{M. J. Way \altaffilmark{1,2},
H. Quintana \altaffilmark{2},
L. Infante \altaffilmark{2},
D. G. Lambas \altaffilmark{3,4}, and
H. Muriel \altaffilmark{3,4}}

\altaffiltext{1}{Space Sciences Division, NASA Ames Research Center, MS 245-6,
Moffett Field, CA 94035.}
\altaffiltext{2}{Department of Astronomy and Astrophysics, Pontificia Universidad
Catolica de Chile, Casilla 306, Santiago 22, Chile.}
\altaffiltext{3}{Grupo de Investigaciones en Astronom\'{\i}a Te\'orica y
Experimental, Observatorio Astron\'omico, Laprida 854, 5000
C\'ordoba, Argentina.}
\altaffiltext{4}{CONICET, Avenida Rivadavia 1917, CP C1033AAJ Buenos Aires, Argentina.} 

\date{Received 2005 January 3 ; accepted 2005 June 24}

\begin{abstract}
The Southern Abell Redshift Survey (SARS) contains 39 clusters of galaxies
with redshifts in the range 0.0 $< z <$ 0.31 and a median redshift
depth of $\bar{z}$ = 0.0845.  SARS covers the region
$0\degr < \delta < -65\degr$, $\alpha < 5^h$, $\alpha > 21^h$
(while avoiding the LMC and SMC) with $\mid b \mid >$ 40.
Cluster locations were chosen from the Abell and Abell-Corwin-Olowin
catalogs while galaxy positions were selected from the Automatic Plate
Measuring Facility galaxy catalog with extinction-corrected magnitudes
in the range 15 $\le$ b$_{j}$ $<$ 19. SARS utilized the Las Campanas
2.5 m duPont telescope, observing either 65 or 128 objects concurrently over
a 1.5 deg$^{2}$ field. New redshifts for 3440 galaxies are reported
in the fields of these 39 clusters of galaxies.
\end{abstract}



\section{INTRODUCTION}

Clusters of galaxies appear as nodes in a gravitationally unfolding
web of dark matter, dark energy and baryons that represents the
large-scale structure of our universe. These nodes are, in fact,
still forming in the present epoch. Therefore, local observations of
clusters of galaxies in comparison with those at higher redshift
\citep[e.g.,][]{Castillo03} in the context of large
cosmological simulations \citep[e.g.,][]{Kochanek03} may in principle lead
to a better understanding of the physical conditions in the early universe.

In order to characterize the clustering of galaxies on multiple scales
many statistical techniques have been used over the years. For example,
using strictly position and velocity information as contained in the
Southern Abell Redshift Survey (SARS), some in wide use are
the cross-correlation function $\xi(r)$ \citep{P80}, the Lee statistic
\citep{F88}, the $\Delta$-test \citep{DS88}, the $\alpha$-test \citep{WB90},
the $\epsilon$-test \citep{B94}, minimal spanning trees \citep{DD04}, extended
friends-of-friends algorithms \citep{Botzler04}, percolation algorithms \citep{Adami02},
and even Minkowski functionals \citep{ShGo99} to name but a few.
However, galaxy velocities in clusters have been traditionally hard to obtain
when one wants to reach volumes and numbers large enough to study a
statistically representative sample of the universe.  To contribute toward
this need, redshift data in the fields of 39 clusters of galaxies have been
collected during several observing sessions over 3 years to carry out
dynamical and morphological analyses. 

The SARS cluster positions were selected from the Abell \citep{A58} and
Abell-Corwin-Olowin \citep[ACO;][]{ACO89} catalogs. Galaxy positions were
taken from the Automatic Plate Measuring (APM) b$_{j}$ catalog \citep{M90a,M90b,M96}.

For a discussion of the velocity dispersions and cluster dynamics in
these clusters see Paper II \citep{M2002}. For information related
to the individual galaxy luminosity profiles in SARS see
\citet{Coenda05a,Coenda05b}

Cluster selection is discussed in $\S$ 2,
galaxy selection from the APM in $\S$ 3, spectral
data observations in $\S$ 4, spectral reductions and velocity data
results in $\S$ 5, survey completeness in $\S$ 6, and cluster identification
in $\S$ 7, and we summarize in $\S$ 8.

\section{THE CLUSTER SAMPLE}

Clusters of richness R$\ge$0 were selected\footnote{The statement by
\citet{M2002} that only Abell clusters of R$>$0 are included in the sample
is incorrect.} from the Abell and ACO catalogs in
the region $0\degr < \delta < -65\degr$, $\alpha < 5^h$, $\alpha > 21^h$
(avoiding the LMC and SMC), with $\mid b \mid >$ 40, in the distance range
represented by Abell distance classes 4 and 5, so that most of the clusters
were covered by the duPont Telescope fiber spectrograph 1.5 deg$^{2}$ field.
Our program was designed to observe a complete sample in
a defined volume near the south Galactic pole while observing
clusters within a reasonable spread of right ascension and declination for observing
accessibility from telescopes in Chile. Our original observing goals were not met for
two reasons: (1) our program was not granted sufficient time at
the Las Campanas Observatory (LCO) 2.5 m telescope; and (2) our original program
was meant to include the use of the 4 m telescope at Cerro Tololo Inter-American Observatory (CTIO),
and we were unsuccessful in obtaining those observations.
Therefore, SARS is not a complete sample of the clusters of galaxies in
the sample volume. 

One of the original goals of the project was to measure the
two point correlation function $\xi(r)$ of clusters of galaxies.
The estimate of $\xi (r)$ depends strongly on edge corrections. If one selects
the objects just by the Abell distance class, the distant (higher redshift)
edge of the cone will have an irregular shape and be ill defined because the
spread of the distance
distributions for D=5 and 6 are wide and mixed \citep{QW90}. To avoid
this problem the distance was estimated by using a combination of magnitudes
and distance class to a depth of z $\approx$ 0.17. The final
determination of the near and far distance limit was to be
precisely defined after the galaxy redshifts were acquired.

To have the full Abell diameter area within the field of view of the two
telescopes originally proposed for use in this project (the CTIO 4 m and the
LCO 2.5 m) we set the low-redshift side of the volume to be
15,000 km s$^{-1}$. This choice also avoids the small number statistics
in cluster density on the low-redshift side of the survey volume cone.

The selected volume of the universe was also chosen to avoid problems
with Milky Way obscuration and differences between north and south.
The observed sample represents a nearly random selection of clusters
in the defined area with a rather even distribution in right ascension, as needed for
telescope observing accessibility (see Fig. 1).

\clearpage
\begin{figure}
\figurenum{1}
\plottwo{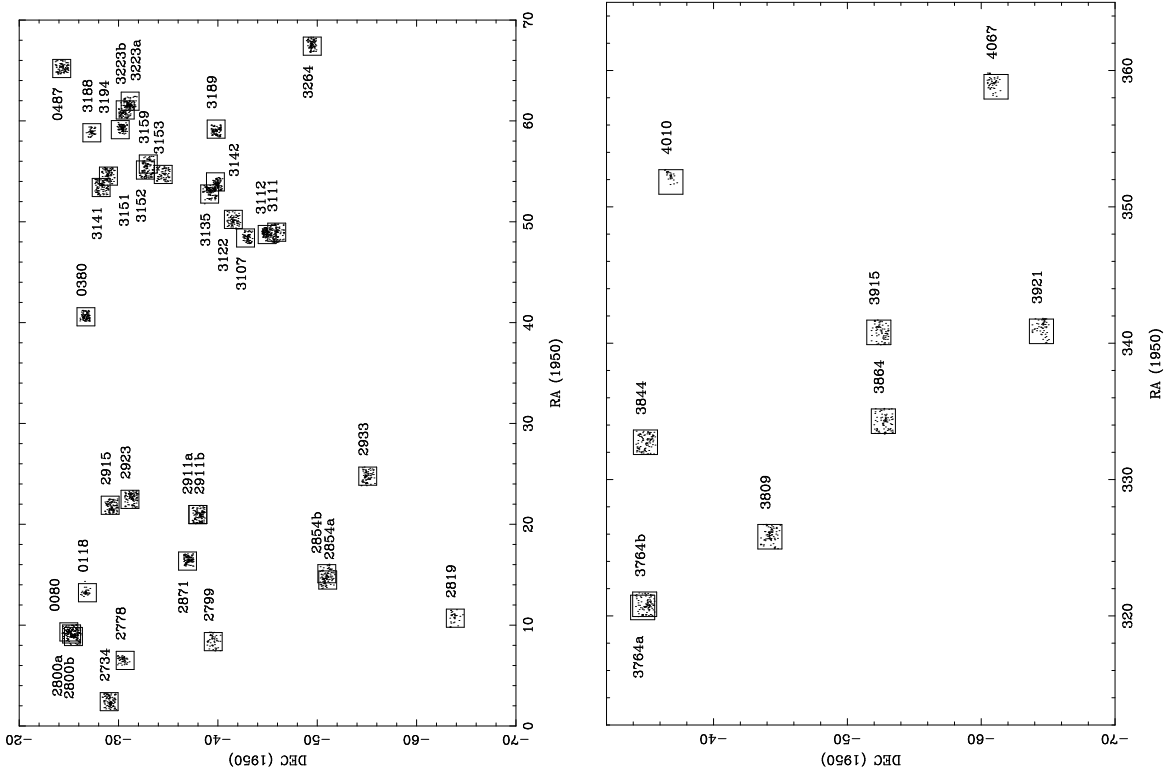}{f1b.ps}
\caption{Measured galaxy velocities with plate centers and 1.5 deg$^{2}$ squares
overlaid for all clusters.
Top: R.A. $=$ 0h to 4$^{h}$40$^{m}$, decl. --20$\deg$ to --70$\deg$ (B1950). Bottom: R.A. $=$
21$^{h}$20$^{m}$ to 24$^{h}$, decl. --32$\deg$ to --70$\deg$ (B1950).}
\end{figure}
\clearpage

The catalog can be shown to be broadly representative of the Abell and
ACO catalogs in a number of ways. The catalog contains a broad
range of Bautz-Morgan types (column 5, Table 1), although it is slightly
biased toward high classes. Also, the regular/irregular types
(column 4, Table 1) as defined by Abell are represented.
The catalog also attempts to provide redshifts for galaxies in moderately rich
clusters (column 7, Table 1), not only in the very richest systems.
This makes SARS useful for studying the properties of clusters corresponding
to a wide range of masses in a nearby sample.

Table 1 lists the information for each observation with the
fiber spectrograph. Column (1) gives the Abell number, columns (2) and (3) right
ascension and declination, column (4) the Abell classification, column (5)
the Bautz-Morgan type, column (6) the Abell number counts, column (7) the Abell
richness class, column (8) the Abell distance class, column (9) the magnitude of
the tenth brightest red cluster galaxy (from the Abell and ACO catalogs),
column (10) the date of the observations and column (11) the final
number of galaxies per cluster field (stars and duplicate measurements
are removed).

\section{THE APM GALAXY SAMPLE}

The galaxies were selected from the APM b$_{j}$ catalog \citep{M90a,M90b,M96}.
The extended version of the APM galaxy catalog contains over 5 million
galaxies brighter than b$_{j}$ = 20.5 over the north and south Galactic
poles covering nearly 10$^{4}$ deg$^{2}$. It was
based on 390 contiguous UK Schmidt Telescope J survey plates. The image surface
brightness profiles were used to distinguish galaxies from stars and merged
images.  Extensive visual checks were performed to quantify the completeness
of the galaxy sample and contamination from stellar images.  Galaxies
in the SARS sample were selected in the range 15 $\le$ b$_{j}$ $<$ 19 and were within
1.5 deg$^{2}$ of the cluster center.\footnote{\citet{M2002} state that
the SARS sample only included galaxies brighter than m$_{R}$=19. 
This is mistaken, as the original APM survey is in b$_{j}$ and galaxies were selected
in the range 15 $\le$ b$_{j}$ $<$ 19.} This is the field
of view of the LCO 2.5 m fiber spectrograph.
Galaxies were also always chosen within 1 Abell radius of the center.
For the first observing session, galaxies were chosen in order of
decreasing APM magnitude such that a significant number over
the number of available fibers was available for selection.
This was needed since the software that allocates fibers to galaxies
discards targets that are within $\sim$30$\arcsec$ of another target
or are too close to the center of the plate (where there is a centering hole).
This strategy obviously produced too many galaxy targets belonging to groups
in the foreground of the cluster. In trying to optimize the use of the
fibers in subsequent telescope sessions we decided to identify
the brightest cluster member and chose galaxies fainter than that.
However, to check on the procedure we choose a few brighter galaxies
that should have been foreground galaxies. A secondary criteria for target
selection was to avoid objects whose images seemed to be of very low
surface brightness since the resulting spectrum would have insufficient
signal-to-noise ratio (S/N) to be useful.  All these procedures allowed
optimal use of the 65 (and later 128) fibers used to collect the spectra.
The APM positions are listed in table 2 and are internally accurate to
within 0$\arcsec$.5.

\section{SPECTRAL OBSERVATIONS}

Spectroscopic observations were carried out with the 2.5 m duPont
telescope of LCO, Chile.  The original multifiber
spectrograph \citep{SH89,SH93} consisted of a plug plate
at the focal plane to which 65 fibers were attached and run to a Boller and
Chivens spectrograph coupled to a 2D-Frutti instrument. 
The 2D-Frutti instrument has a blue-sensitive Carnegie image tube as the
first stage followed by a middle stage consisting of a series of microchannel
plates ending with a continuously read rocking CCD.  A 600 line mm$^{-1}$
grating blazed at 5000{\AA} was set at an angle of $9^{\circ}$  40$\arcmin$, giving
wavelength coverage from $\sim$3800 to 6800 {\AA}. Normally, 50--55 fibers were used
for objects. Ten sky fibers were set aside, spaced at intervals of one
every six fibers along the spectrograph entrance, and positioned in a
random pattern in the plug plate. The resulting 2D-Frutti image had a 1520 x
1024 pixel area, with a dispersion of $\sim$2.6 {\AA} pixel$^{-1}$ and
a final resolution of $\sim$10 {\AA}.  The fiber images were $\sim$8 pixels
wide and separated by $\sim$12 pixels from center to center.

By the time of the 1992 January observations, the Boller and Chivens
spectrograph had been replaced by the ``floor fiber spectrograph" built
by Steve Shectman. This had 128 fibers versus the original 65
but with the same wavelength coverage and resolution.
At least 12 fibers out of 128 were normally dedicated to sky.
The rest of the setup was the same as for the 65 fiber instrument.

As mentioned above, the front detector of the 2D-Frutti instrument was a
blue-sensitive Carnegie image tube whose output was imaged on a stack of
microchannel plates. The complexities in these intermediate microchannel plate
stages introduced significant position-dependent pixel-to-pixel variations,
but these were independent of the object wavelength. Since no
calibration in flux was intended, any wavelength-dependent pixel-to-pixel
variations of the front unit were not significant when
compared to the noise introduced by the pixel-to-pixel variations of
the intermediate section. Hence, quartz lamp exposures were used to
correct for pixel to pixel variations of the detector (also known as
a ``flat field''). To properly illuminate the entire detector surface, and
consequently the following microchannel stages, the grating angle
was changed to several values during these exposures.

Helium-neon comparison lamp exposures were taken off the
wind screen for wavelength calibration before and after each exposure. The
2D-Frutti detector has a small dark current, and hence no corrections were made for
that effect.

Exposure times were adjusted to be between 60 and 148 minutes, depending on the
brightness of the selected galaxies for each exposure. The 2D-Frutti instrument
is a photon-counting system with which one can view the current exposure at any
time. In this way one can obtain the optimum exposure time for each field.
Between 28 and 196 spectra in each cluster field were obtained. See Table 1
for more details.

\section{REDSHIFT REDUCTIONS}

Radial velocity determinations were carried out using a cross-correlation
technique and when necessary by identifying and fitting by eye absorption-
or emission-line profiles.  All reductions were performed inside
the IRAF \citep{T93} environment.
For a more complete discussion of the reductions, see \citet{QRW96}.
Below is a summary of the reductions.

Due to the nature of the fiber+2D-Frutti system, typical {\bf S}-shaped distortions
are inherent in this instrument.  A sixth-order spline3 \citep{NR1} curve
was used to trace the {\bf S}-shaped spectra.  The IRAF HYDRA package was
used to extract the spectra, correct pixel-to-pixel variations
via the dome flat, use a fiber transmission table for appropriate sky
subtraction and put the spectra on a linear scale in wavelength. The wavelength
solutions for 20--30 points using a fifth, sixth, or seventh order Chebyshev
polynomial \citep{NR2} typically yielded residual rms values less than 0.4 {\AA},
where 1 pixel $\sim$2.6{\AA}. The 10 or more sky spectra from each exposure
were combined via a median filter and subtracted from each of the
object spectra.

Two different methods were used to measure the redshift of the objects.
For most normal early type galaxy spectra the RVSAO \citep{K91}
cross-correlation algorithm supported inside IRAF was used.
The algorithm used in RVSAO is described in \citet[][hereafter TD79]{TD79}.
A redshift reliability/quality factor, the R-value, is generated by RVSAO
(see TD79 for details).  Normally, a low R-value (R$\leq{4}$) indicated
a need to look at the spectra and try line-by-line Gaussian fitting
(the second method). If the line-by-line fitting agreed to within 1 $\sigma$
of the original RVSAO result with the lowest template match error, then
the RVSAO result was chosen (see next paragraph for further explanation).
If it did not, then the radial velocities from each line fit were averaged,
and the standard deviation became the quoted error. If there were observable
emission and absorption lines the fits were taken from the absorption
lines only. All radial velocities had a heliocentric correction applied.

To us RVSAO, four
template spectra set to z = 0 with high S/Ns and well-determined radial
velocities were used. Two of the four templates used in this paper were galaxy
spectra with high S/N, NGC 1407 and NGC 1426, taken with the fiber instrument.
A third template galaxy, NGC 1700, was from the previous detector on the
2.5 m at LCO (formerly known as the Shectograph). The fourth
template was a synthetic spectrum. The synthetic template was constructed
from the excellent library of stellar spectra of \citet{JHC84}. Ratios of
stellar light were used for the E0 galaxy NGC1374 from the synthesis
studies of \citet{PIC85}. In the end it was found that of the four radial velocity
cross-correlation templates mentioned above, the template that
gave the lowest error value and had R$>$4 provided more consistent
results. It was also found that typically only one of the four templates provided
values of R$>$4. The cause of this behavior was related to a couple of
different factors: (1) our galaxy templates were chosen to have slightly different
galaxy types (E0 to E4) in order to cover the possible range of observed
objects, not including AGN-type objects; and (2) the galaxy templates
were observed with different instruments or synthesized. These factors resulted
in different equivalent widths between galaxy templates for
many of the largest absorption line features. Hence, when a template match
with R$>$4 was found the others template matches had typical values
of R$\leq{4}$ and inconsistent values of radial velocity.

Table 2 contains the final velocities for all members in the fields
of the Abell clusters discussed in this paper. Columns (1) and (2) contain
the right ascension and declination, column (3) contains the velocity, and
column (4) contains the velocity error. If the value of column (4) is --99, this
means that only one or two strong emission or absorption lines were found, and no
estimate of the error can be given. Column (5) contains the TD79 reliability
number. When column (5) is empty column (6) contains
the number of individual emission or absorption lines used in the
velocity estimation. Column (7) contains the APM b$_{j}$ magnitude of the galaxy.

\section{SELECTION FUNCTION, REDSHIFT COMPLETENESS AND REDSHIFT COMPARISON}

The 2dF Galaxy Redshift Survey (2dFGRS) has a carefully determined selection
function \citep{N2002}
based on extinction corrected b$_{j}<$19.45 APM magnitudes. This is nearly the
same as SARS, which is limited to 15 $\le$ b$_{j}$ $<$ 19. Since seven of the SARS cluster
fields (A118, A380, A2734, A2778, A2915, A2923, and A3844) overlap the
2dFGRS, the SARS redshift completeness can be evaluated against the 2dFGRS.
The 2dFGRS b$_{j}$ magnitudes and redshifts in the fields of these SARS clusters
were obtained from the Web site interface to the 2dFGRS mSQL database
\footnote{see http://www.mso.anu.edu.au/2dFGRS}.  See \cite[][their \S 9.2]{Colless01}
for details. The relevant 2dFGRS query for this study is:

	SELECT ra, dec, BJSEL, z, quality from public WHERE extnum=0 AND
quality$>$=3 AND BJSEL$<$19.0 

Hence, the 2dFGRS query included only those galaxies with redshifts of quality
better than 3 and b$_{j}$$<$19. They also exactly overlapped a square right
ascension and declination boundary in the fields of the SARS clusters.

All APM b$_{j}$ magnitudes in the same overlapping regions are needed to
properly characterize the SARS selection function. Hence, all the APM b$_{j}$
magnitudes in the fields of the seven SARS clusters were taken from the
SuperCOSMOS/APM Web site interface to the database
\footnote{see http://www-wfau.roe.ac.uk/sss/obj.html} using the following query values:
image quality, 65535; paring radius, 3$\arcsec$; color correct magnitudes, yes,
and b$_{j}<$19.

Figure 2 shows the b$_{j}$ magnitude distribution for the APM (8140 objects for
15 $\le$ b$_{j}$ $<$ 19), 2dFGRS (1274 objects for 15 $\le$ b$_{j}$ $<$ 19), and SARS fields
(596 objects) in the seven overlapping clusters.

Figure 3 demonstrates the redshift completeness as a
function of b$_{j}$ magnitude for the seven fields in the 2dFGRS and SARS
catalogs with respect to the APM catalog ({\it left-hand axis}), and for
SARS with respect to 2dFGRS ({\it right-hand axis}).

\clearpage
\begin{figure}
\figurenum{2}
\plotone{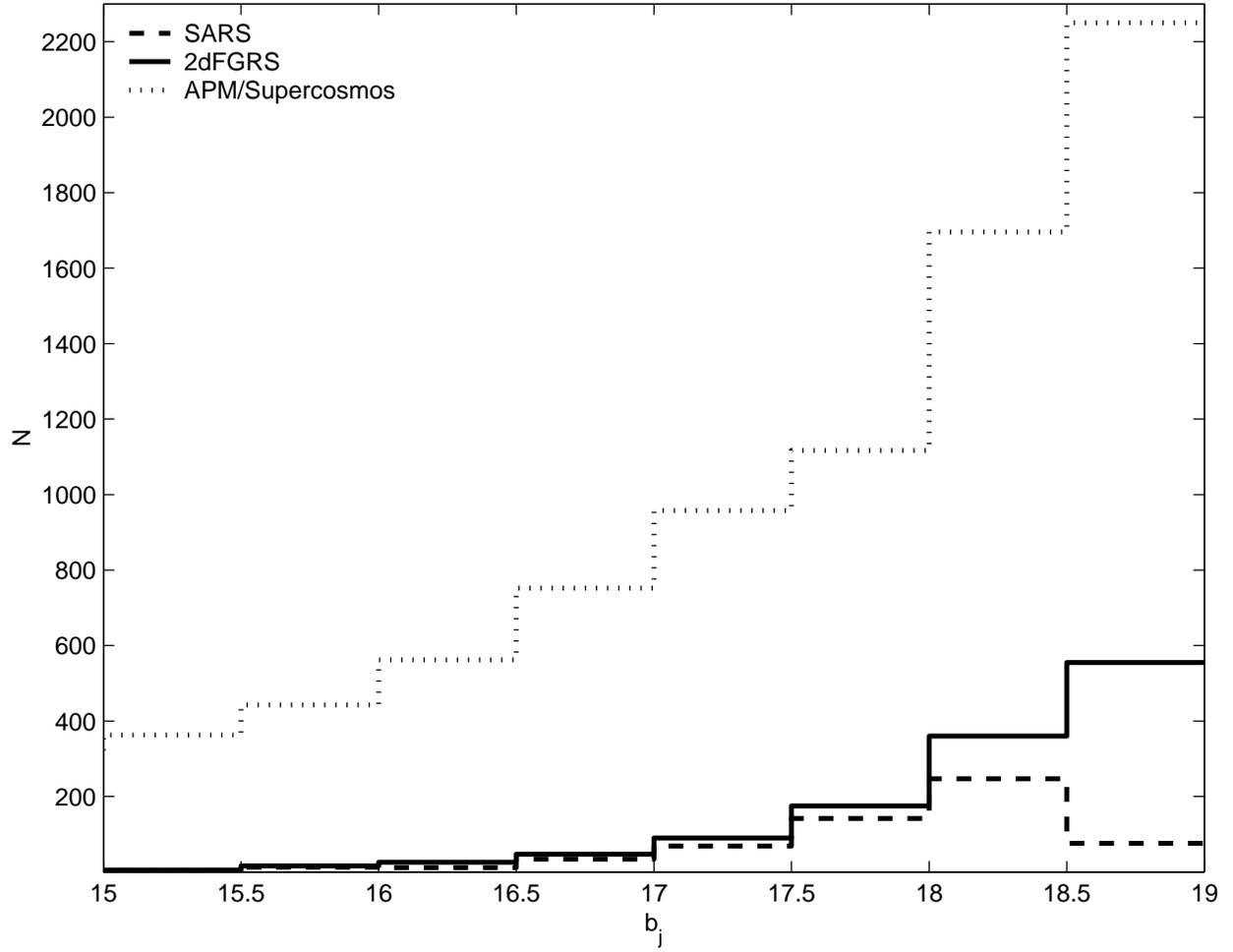}
\caption{Magnitude distribution for the seven overlapping fields in the APM,
2dFGRS, and SARS samples.}
\end{figure}

\begin{figure}
\figurenum{3}
\plotone{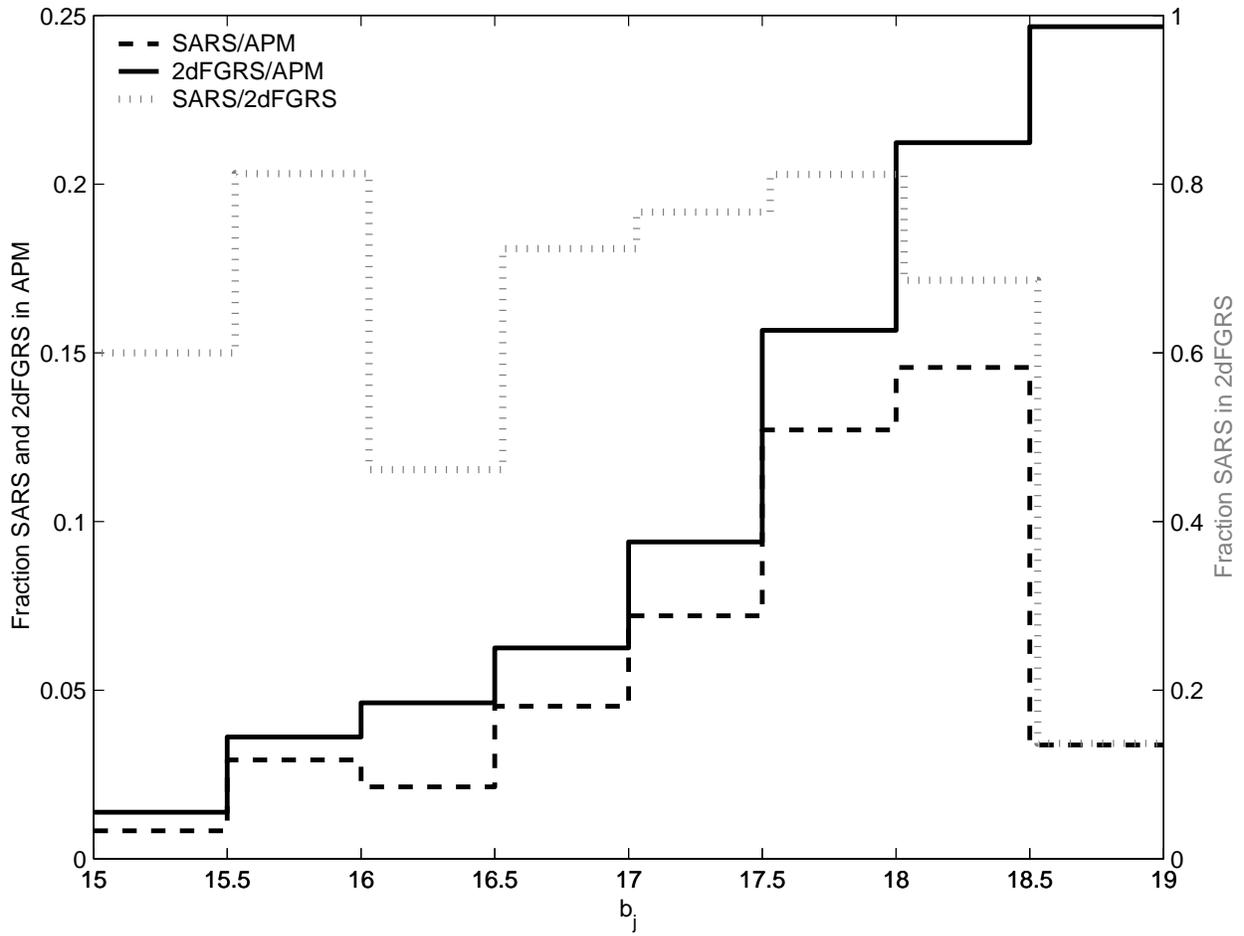}
\caption{Redshift completeness for the seven overlapping fields in the APM, 2dFGRS,
and SARS catalogs.}
\end{figure}
\clearpage

Figure 4 shows the redshift distribution of the entire SARS
survey ({\it gray histogram}), the 2dFGRS redshifts in the seven SARS fields
({\it black histogram}), and redshifts in the seven SARS fields that overlap the 2dFGRS
({\it white histogram}). A simple smooth analytic approximation curve taken from
the 2dFGRS \citep{Colless01} is fitted to the entire SARS survey data and
superposed ({\it gray line}). Here $\bar{z}$ is defined as the median
redshift, dN$\propto$$z^{2}$exp[-(1.36$z/\bar{z}$)$^{1.55}$]dz.

\clearpage
\begin{figure}
\figurenum{4}
\plotone{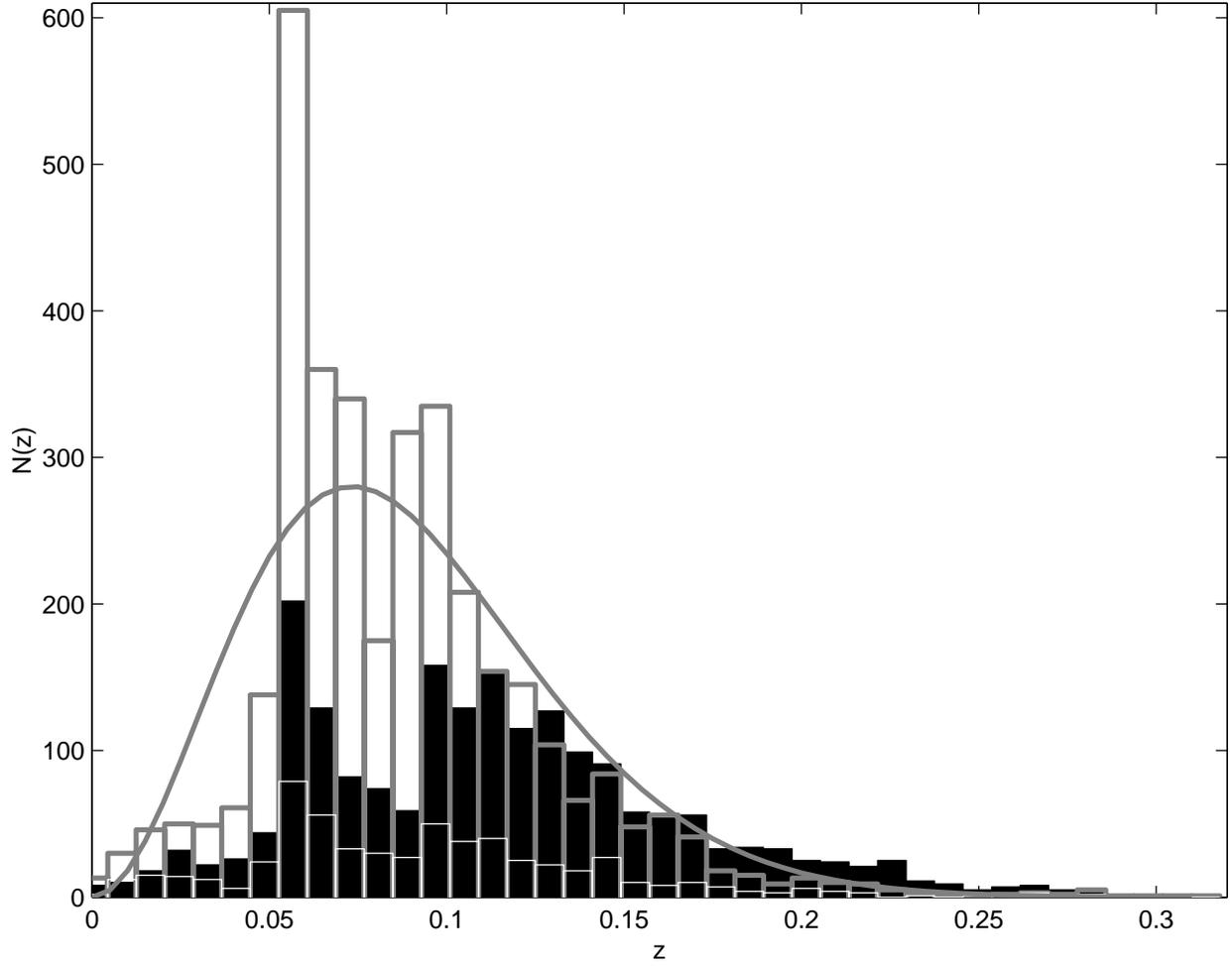}
\caption{Redshift distribution of the entire SARS survey ({\it gray histogram}), the
2dFGRS redshifts in the seven SARS fields ({\it black histogram}),
redshifts in the seven SARS fields that overlap the 2dFGRS ({\it white histogram}),
and an analytical approximation curve ({\it gray curve}) fitted to the entire SARS survey
(see text).}
\end{figure}
\clearpage

It is clear from Figures 2--4 that the 2dFGRS is more complete,
particularly at fainter magnitudes and redshifts, in the fields of the seven SARS
clusters.  Also note that the median redshift of the entire SARS catalog
is $\bar{z}$=0.0845, while the full 2dFGRS has a higher value of $\bar{z}$=0.11.

There are several reasons for these differences. First, the 2dFGRS observed
many overlapping fields to lessen the effect of fiber collisions.
See figure 15 \citep{Colless01} regarding the 2dFGRS redshift completeness
function. This makes the 2dFGRS selection of galaxies in these fields more
complete than SARS since SARS was only able to obtain two overlapping
observations in the fields of five clusters (see Table 1 and Fig. 1), and
none of those are in the seven 2dFGRS overlapping fields.  Second, the 2dFGRS
has 400 fibers spread over a 2$\deg$ diameter field of view, whereas SARS
used only either 65 fibers (2dFGRS overlap fields A118 and A2778) or 128 fibers
(2dFGRS overlap fields A380, A2734, A2915, A2923, and A3844) over a 1.5 deg$^{2}$
diameter field of view. The latter is especially obvious, since the percentage
difference of fibers per unit area of the 65 fiber SARS runs compared to 2dFGRS
is 28.8\%. For the 128 fiber SARS runs the value is 56.8\%.
When averaged over the total number of observed objects in the seven overlapping
fields of SARS (596 objects) and 2dFGRS (1274 objects), one obtains
596/1274=46.8\%, which is what one expects from the previous two percentages
when weighted accordingly.  Third is the fact that the 2dFGRS is a slightly
deeper survey due to its larger aperture (4 m vs. 2.5 m) and better detector
technology (CCD camera vs. a less quantum efficient photon counting system).
This fact can also be seen in Figure 4 as the 2dFGRS redshifts just in the
seven SARS fields (1274 objects) start to outnumber even the entire
SARS survey (3440 objects) at higher redshifts.
Fourth, this could also be due to the more manual selection of
galaxies in the SARS survey (see Section 3 above) which tried to
avoid selecting foreground and background galaxies in the Abell fields.
The avoidance of selecting fainter background galaxies demonstrates itself
in Figures 2 and 3.

In attempting to compare the same SARS and 2dFGRS redshifts in the seven
overlapping fields, one finds 530 objects that reside within 5$\arcsec$ of each
other in right ascension and declination. After applying an iterative 3 $\sigma$
clipping to the redshift differences of these 530 overlapping objects, 444 are left
with an average redshift difference of SARS minus the 2dFGRS of
+39.3 km s$^{-1}$, with a standard deviation of 138 km s$^{-1}$.

\section{Cluster identification}

Several authors \citep{VH97} have discussed the consequences of projection
effects when clusters are selected in two dimensional catalogs. Redshift surveys provide
precise information on the reality of the clusters selected. Figure 5 shows the
ensemble of velocities for each cluster. In several cases, the presence of a
superposition of two clusters along the line of sight is obvious: A380, A2871,
A2911, A3135, A3142, A3152 and A3864. Moreover, in many other cases
there are more than two concentrations along the line of sight.  For example,
there are three large concentrations in A2915, A3122, A3223, A3864 and A3844
including many smaller concentrations.  For a more in depth discussion
of cluster identification, see Paper II.

\clearpage
\begin{figure}
\figurenum{5}
\epsscale{0.9}
\plotone{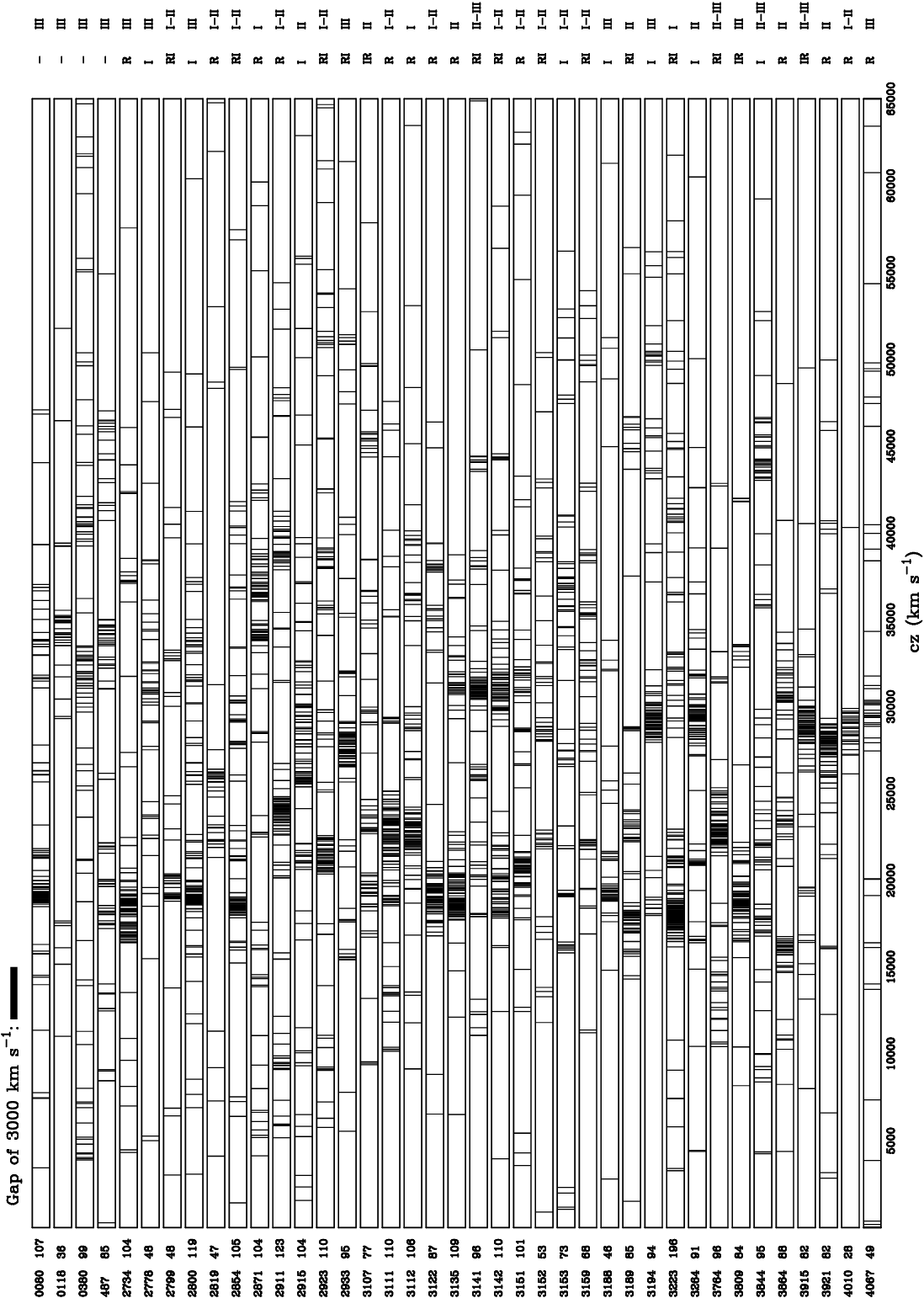}
\caption{Redshift distribution for the catalog.}
\end{figure}
\clearpage

\section{Summary}

We presented a new redshift catalog of 3440 galaxies in the fields of
39 southern Abell clusters of galaxies selected from the
Abell and ACO catalogs. The redshifts and corresponding
APM magnitudes presented should provide a means to better characterize
clusters of galaxies in the local universe when combined with the largest
redshift surveys available today, the 2dFGRS \citep{Colless01,Colless03} and the SDSS
\citep{York2000}. They should also be helpful in characterizing clusters
of galaxies in large volume photometric surveys of the universe
\citep{Bahcall03} by increasing the number of densely mapped and carefully
studied clusters of galaxies. 


\acknowledgements
M.J.W. acknowledges support from the NASA-Ames Research Center Director's
Discretionary Fund.  H.Q. and L.I. acknowledge the FONDAP
Centro de Astrofisica project, Chile, for partial support.
H.Q. is grateful for the award of a
Guggenheim Fellowship. This work was partially supported by the Consejo de 
Investigaciones Cient\'{\i}ficas y T\'ecnicas de la Rep\'ublica Argentina, 
CONICET, the Agencia C\'ordoba Ciencia, C\'ordoba, the 
Secretaria de Ciencia y T\'ecnica, UNC, the Agencia Nacional de Promoci\'on 
Cient\'ifica and the Fundaci\'on Antorchas, Argentina.
The authors acknowledge the use of SuperCOSMOS Sky Survey material,
which is based on photographic data originating from the UK, Palomar,
and ESO Schmidt telescopes and is provided by the Wide-Field Astronomy
Unit, Institute for Astronomy, University of Edinburgh.
The authors also acknowledge the use of the 2dF Galaxy Redshift Survey database.
The 2dF Galaxy Redshift Survey has been made possible by the dedicated efforts
of the staff of the Anglo-Australian Observatory, both in creating the
2dF instrument and in supporting it on the telescope. The Anglo-Australian
Observatory is funded by the Australian government (through DEST) and
the UK government (through PPARC).

\clearpage

\begin{table}
\footnotesize
\begin{center}
\caption{Cluster Observations - Spectroscopy}
\begin{tabular}{ccccccccccc}
\hline \hline
      &  R.A.  & Decl.     &  &      &     & Richness & Distance  &  & Date      & \\
Abell &(B1950.0)&(B1950.0) &Type\tablenotemark{a}& Bautz-Morgan&Counts\tablenotemark{b}\
& Class    & Class     &m$_{10}$\tablenotemark{c} & Observed  & Number\tablenotemark{d} \\
(1)  &  (2)    &    (3)    &(4)&(5) & (6) & (7) & (8)  & (9) & (10) & (11) \\
\hline
0080 & 0.63167 & -24.95000 & I & III & 64  & 0 & 4 & 15.8 & 1993/10 &  107 \\
0118 & 0.88833 & -26.68333 & I & II-III & 77  & 1 & 5 & 17.2 & 1991/09 &   36 \\
0380 & 2.70333 & -26.46667 & RI & II & 64  & 2 & 5 & 17.2 & 1991/11 &   99 \\
0487 & 4.34603 & -24.28309 & I & II-III & 104 & 2 & 5 & 17.2 & 1992/01 &   85 \\
2734 & 0.14667 & -29.15000 & R & III & 58 &  1 & 4 & 16.3 & 1993/10 &  104 \\
2778 & 0.43500 & -30.51667 & I & III & 51 &  1 & 5 & 17.0 & 1991/09 &   48 \\
2799 & 0.58500 & -39.40000 & RI & I-II & 63 &1 & 4 & 16.2 & 1991/09 &   48 \\
2800 & 0.59167 & -25.36667 & I & III & 59 &  1 & 4 & 15.8 & 1991/09,1993/09 &  119 \\
2819 & 0.72833 & -63.86667 & R & I-II & 90 & 2 & 4 & 16.0 & 1991/09 &   47 \\
2854 & 0.97667 & -50.80000 & RI & I-II & 64& 1 & 4 & 15.8 & 1991/09,1993/09 &  105 \\
2871 & 1.09333 & -37.00000 & R & I    & 92 & 2 & 5 & 16.7 & 1991/11 &  104 \\
2911 & 1.39667 & -38.23333 & R & I-II & 72 & 1 & 4 & 16.3 & 1991/09,1993/09 &  123 \\
2915 & 1.44167 & -29.26667 & I & II & 55  &  1 & 5 & 16.9 & 1991/11 &  104 \\
2923 & 1.50000 & -31.35000 & RI & I-II & 50 &1 & 5 & 17.0 & 1991/11 &  110 \\
2933 & 1.64667 & -54.81667 & RI & III & 77 & 1 & 5 & 16.9 & 1993/10 &   95 \\
3107 & 3.22667 & -42.95000 & IR & II & 61 &  1 & 5 & 17.0 & 1992/12 &   77 \\
3111 & 3.26833 & -45.91667 & R & I-II & 54 & 1 & 4 & 16.3 & 1991/11 &  110 \\
3112 & 3.27000 & -44.41667 & R & I & 116 &   2 & 4 & 16.1 & 1992/01 &  106 \\
3122 & 3.34167 & -41.51667 & R & I-II & 100 &2 & 4 & 15.8 & 1992/01 &   87 \\
3135 & 3.53667 & -39.16667 & R & II & 111 &  2 & 3 & 15.5 & 1991/11 &  109 \\
3141 & 3.58167 & -28.21667 & RI & II-III & 55&1& 5 & 16.8 & 1992/01 &   96 \\
3142 & 3.58167 & -39.96667 & RI & I-II & 78 & 1& 5 & 16.9 & 1991/11 &  110 \\
3151 & 3.64000 & -28.86667 & R & I-II & 52 & 1 & 4 & 16.0 & 1991/11 &  101 \\
3152 & 3.64000 & -32.73333 & RI & I-II & 51 & 1& 5 & 17.0 & 1992/01 &   53 \\
3153 & 3.65167 & -34.41667 & I & I-II & 64 & 1 & 5 & 17.0 & 1992/12 &   73 \\
3159 & 3.70167 & -32.85000 & RI & I-II & 98 & 2 &5 & 17.1 & 1992/12 &   68 \\
3188 & 3.92833 & -27.18333 & I & III & 67 & 1 &  5 & 17.1 & 1992/12 &   46 \\
3189 & 3.93333 & -39.75000 & RI & II & 65 &  1 & 6 & 17.4 & 1992/12 &   85 \\
3194 & 3.95333 & -30.31667 & I & III  & 83 & 2 & 5 & 16.7 & 1992/01 &   94 \\
3223 & 4.11000 & -30.95000 & RI & I  & 100 & 2 & 3 & 15.6 & 1991/11,1992/12 &  196 \\
3264 & 4.50167 & -49.41667 & I & II  & 53  & 1 & 5 & 16.7 & 1992/01 &   91 \\
3764 & 21.38000 & -34.93333 & RI &II-III &53 &1 & 5 & 16.8 & 1993/05,1991/09 &  96 \\
3809 & 21.73000 & -44.13333 & IR & III & 73 & 1 & 4 & 16.0 & 1993/05 &  84 \\
3844 & 22.17667 & -35.00000 & I & II-III&52 & 1 & 5 & 17.0 & 1993/10 &  95 \\
3864 & 22.28333 & -52.73333 & R & II   & 60 & 1 & 5 & 16.7 & 1993/09 &  88 \\
3915 & 22.74333 & -52.31667 & IR &II-III& 55 &1 & 5 & 17.0 & 1993/09 &  82 \\
3921 & 22.77500 & -64.65000 & R & II   & 93 & 2 & 5 & 16.9 & 1993/09 &  82 \\
4010 & 23.47500 & -36.78333 & R & I-II & 67 & 1 & 5 & 16.9 & 1991/09 &  28 \\
4067 & 23.94000 & -60.95000 & R & III  & 72 & 1 & 5 & 17.1 & 1991/09 &  49 \\
\hline \hline
\end{tabular}
\tablenotetext{a}{cluster classification in Abell's system: I = irregular,
R = regular, IR and RI = intermediate}
\tablenotetext{b}{number of cluster members between m3 and m3+2, corrected
for background \citep{A58, ACO89}}
\tablenotetext{c}{red magnitude of the tenth brightest cluster member
\citep{A58, ACO89}}
\tablenotetext{d}{number of galaxy velocities measured in field of cluster}

\end{center}
\end{table}

\begin{table}
\footnotesize
\caption{Velocity Data\tablenotemark{c}}
\begin{tabular}{ccccccc}
\hline \hline
RA     & Dec    & Velocity       & error          &               & & $m_{APM}$\\
(B1950.0) & (B1950.0) & (km~ s$^{-1}$) & (km~ s$^{-1}$) &R\tablenotemark{a}&N\tablenotemark{b}& (b$_{j}$)\\
(1) & (2) & (3) & (4) &(5)&(6)& (7)\\
\hline
Abell 0080 \\
  8.54304 & -24.89661 & 19134 & 89 &       &      4 & 17.9 \\
  8.59467 & -25.72231 & 18456 & 91 &  4.02 &        & 17.3 \\
  8.62092 & -24.90189 & 19466 & 63 &  5.26 &        & 18.1 \\
  8.62167 & -24.88022 & 19090 & 19 & 17.38 &        & 16.4 \\
  8.67258 & -24.43053 & 26299 & 68 &       &      4 & 18.1 \\
  8.69967 & -24.77514 & 26740 & 40 &  6.67 &        & 17.3 \\
  8.72821 & -25.44467 & 18757 & 32 &  8.40 &        & 17.0 \\
  8.76542 & -25.68603 & 18848 & 32 &  8.14 &        & 16.7 \\
  8.76738 & -25.64111 & 11335 & 81 &  2.17 &        & 18.2 \\
  8.77308 & -25.48561 & 19327 & 24 & 14.41 &        & 18.2 \\
  8.78842 & -25.23067 & 18785 & 20 & 15.55 &        & 17.7 \\
  8.80158 & -25.06683 & 46824 & 55 &  4.17 &        & 18.1 \\
  8.82975 & -25.27392 & 19610 & 44 &  7.05 &        & 16.8 \\
  8.84508 & -25.41200 & 18840 & 22 & 14.45 &        & 16.5 \\
  8.85296 & -25.16981 & 19953 & 28 & 10.78 &        & 17.5 \\
  8.86013 & -24.41097 & 15754 & 35 &  8.16 &        & 17.0 \\
  8.86413 & -24.82242 & 20200 & 79 &  2.70 &        & 18.1 \\
  8.87579 & -24.45431 & 33853 & 59 &  6.89 &        & 18.2 \\
  8.88133 & -25.35728 & 18862 & 17 & 20.69 &        & 16.8 \\
  8.88138 & -25.10375 & 19086 & 17 & 18.28 &        & 17.3 \\
  8.88567 & -24.85181 & 19223 & 87 &  4.13 &        & 17.9 \\
  8.90158 & -25.55344 & 18812 & 43 &  7.46 &        & 18.1 \\
  8.90346 & -25.52053 & 18720 & 32 &  9.09 &        & 16.6 \\
  8.91017 & -25.34245 & 19952 & 15 & 23.86 &        & 15.8 \\
  8.91100 & -25.46392 & 19132 & 25 & 12.23 &        & 16.9 \\
  8.91158 & -25.36586 & 23155 & 90 &  2.74 &        & 17.9 \\
  8.92204 & -25.61689 & 19015 & 24 & 13.30 &        & 16.8 \\
  8.92288 & -25.50572 & 18698 & 18 & 18.29 &        & 17.3 \\
  8.92817 & -25.71933 & 19567 & 72 &  2.75 &        & 18.1 \\
  8.95354 & -25.16981 & 33568 & 29 & 10.65 &        & 18.0 \\
  8.97571 & -24.70645 & 26275 & 24 & 13.50 &        & 16.9 \\
  8.99808 & -25.27289 & 19571 & 28 & 10.98 &        & 17.8 \\
  9.01488 & -25.27667 & 18896 & 43 &  6.40 &        & 18.0 \\
  9.02279 & -25.19039 & 19145 & 24 & 12.62 &        & 16.7 \\
  9.03167 & -25.62456 & 19313 & 81 &  2.99 &        & 17.4 \\
  9.03367 & -25.48383 & 19255 & 84 &  3.19 &        & 18.1 \\
  9.03679 & -25.46317 & 33824 & 28 & 12.10 &        & 17.9 \\
  9.04567 & -25.60708 & 18991 & 21 & 15.15 &        & 17.8 \\
  9.07358 & -24.84483 & 25640 & 32 &  8.38 &        & 18.1 \\
  9.07750 & -25.11583 & 19128 & 41 &  6.80 &        & 17.9 \\
  9.09667 & -25.33556 & 19214 & 30 &  9.96 &        & 16.7 \\
  9.13733 & -24.64961 & 21245 & 25 & 12.44 &        & 17.0 \\
  9.14379 & -25.55219 & 19023 & 34 &  8.27 &        & 17.7 \\
\hline \hline
\end{tabular}
\tablenotetext{a}{Redshift reliability number, see text for details.}
\tablenotetext{b}{Number of emission or absorption lines measured, see text
for details.}
\tablenotetext{c}{Table 2 is published in its entirety in the electronic edition
of the Astronomical Journal. A portion is shown here for guidance regarding its
form and content. Table 2 is available in its entirety via the link to the machine--readable
version above.}
\end{table}

\clearpage




\end{document}